\documentclass[]{spie}  

\usepackage{amsmath,amsfonts,amssymb}
\usepackage{graphicx}
\usepackage[multi-part-units=single]{siunitx}
\usepackage{color}
\usepackage{bm}
\usepackage{url}
\usepackage[export]{adjustbox}
\usepackage{multirow}
\usepackage[colorlinks=true, allcolors=blue]{hyperref}

\title{Spatio-spectral deep learning methods for \emph{\bfseries{\fontsize{15.88pt}{100}\selectfont{in-vivo}}} hyperspectral laryngeal cancer detection}

\author[a]{Marcel Bengs$^*$}
\author[b]{Stephan Westermann$^*$}
\author[a]{Nils Gessert}
\author[d]{Dennis Eggert}
\author[c]{Andreas O. H. Gerstner}
\author[b]{Nina A. Mueller}
\author[d]{Christian Betz}
\author[b,d]{Wiebke Laffers}
\author[a]{Alexander Schlaefer}
\affil[a]{Institute of Medical Technology, Hamburg University of Technology, Am
Schwarzenberg-Campus 3, Hamburg 21073, Germany;} 
\affil[b]{Department of Otorhinolaryngology/Head and Neck Surgery, University of Bonn, Sigmund-Freud-Str. 25, Bonn 53127, Germany;} 
\affil[c]{Klinikum Braunschweig, ENT-Clinic, Holwedestr. 16, Braunschweig 38118, Germany;} 
\affil[d]{University Medical Center Hamburg-Eppendorf, Clinic and Polyclinic for Otolaryngology, Martinistrasse 52, Hamburg 20246, Germany}

\authorinfo{Further author information: (Send correspondence to Marcel Bengs) Marcel Bengs: E-mail: marcel.bengs@tuhh.de
\\ $^*$ Both authors contributed equally.}

\begin{document} 
\maketitle 
\fontsize{10}{10}\selectfont
\begin{abstract}

Early detection of head and neck tumors is crucial for patient survival. Often, diagnoses are made based on endoscopic examination of the larynx followed by biopsy and histological analysis, leading to a high inter-observer variability due to subjective assessment. In this regard, early non-invasive diagnostics independent of the clinician would be a valuable tool. A recent study has shown that hyperspectral imaging (HSI) can be used for non-invasive detection of head and neck tumors, as precancerous or cancerous lesions show specific spectral signatures that distinguish them from healthy tissue. However, HSI data processing is challenging due to high spectral variations, various image interferences, and the high dimensionality of the data. Therefore, performance of automatic HSI analysis has been limited and so far, mostly \textit{ex-vivo} studies have been presented with deep learning. In this work, we analyze deep learning techniques for \textit{in-vivo} hyperspectral laryngeal cancer detection. For this purpose we design and evaluate convolutional neural networks (CNNs) with 2D spatial or 3D spatio-spectral convolutions combined with a state-of-the-art Densenet architecture. For evaluation, we use an \textit{in-vivo} data set with HSI of the oral cavity or oropharynx. Overall, we present multiple deep learning techniques for \textit{in-vivo} laryngeal cancer detection based on HSI and we show that jointly learning from the spatial and spectral domain improves classification accuracy notably. Our 3D spatio-spectral Densenet achieves an average accuracy of 81\%. 
\end{abstract}

\keywords{Hyperspectral imaging, convolutional neural networks, optical biopsy, intraoperative imaging, head and neck cancer}

\section{INTRODUCTION}
Early detection of head and neck cancer is essential for a patient's prognosis \cite{horowitz2001perform}. Especially, laryngeal cancer has a high incidence, because the malignancy of lesions is often detected too late \cite{habermann2001carcinoma, allison1998predictors}. Typically, diagnosis of malignant tumors in the head and neck area are made based on endoscopic examination of the larynx, followed by an invasive biopsy and histological examination, which is considered to be the gold standard. However, invasive biopsies can lead to significant functional deterioration \cite{alieva2018potential}. Moreover, the prognostic quality of the diagnosis and histological examination are strongly dependent on the experience of the clinician, thus the applied examination methods are neither objective nor quantitative \cite{kujan2007oral, khalid2009reinterpretation}. Consequently, a non-invasive diagnostic method independent of the clinician would be an important milestone for the detection of laryngeal cancer, greatly increasing the chance of a successful treatment. 

A recent study has shown that the diagnosis of suspicious mucosal lesions can be significantly improved by means of extended diagnostics using optical instruments \cite{lohler2014incidence}. The application of optical instruments for oral cancer detection is based on the fundamental assumption that tissue characteristics change during a disease (e.g. due to angiogenesis, increased blood circulation or inflammation) and that there are deviating absorption, scattering, fluorescence, and metabolic properties compared to healthy tissue \cite{lu2014medical}. In this regard, narrow band imaging (NBI) in combination with induced fluorescence \cite{arens2007indirect} or OCT \cite{volgger2013evaluation} have been used to analyze tumor boundaries and distinguish benign from malignant tissue. 

Moreover, hyperspectral imaging (HSI) has shown promising results for laryngeal cancer detection \cite{regeling2016hyperspectral}. HSI is a noncontact imaging modality that acquires a series of images at several spectral bands, generating a hyperspectral image cube \cite{signoroni2019deep}. Considering HSI for \textit{in-vivo} laryngeal cancer detection, previous work has focused on the development of an image pre-processor for HSI \cite{regeling2016development}. Moreover, a previous study has shown the feasibility of detecting laryngeal cancer based on \textit{in-vivo} HSI, by using traditional machine learning algorithms \cite{laffers2016early}. However, analyzing HSI is a challenging task, because of high spectral variations, the high dimensionality of the data and the presence of redundancy due to the high spectral resolution\cite{signoroni2019deep}.

Recently, deep learning techniques have shown promising performance for HSI with applications ranging from agriculture, food quality to biomedicine \cite{signoroni2019deep}. In particular, previous studies evaluated convolutional neural networks (CNNs) for classifying head and neck cancer based on HSI of excised tissue samples, demonstrating the potential of deep learning as a reliable \textit{ex-vivo} classifier. However, an \textit{in-vivo} evaluation is still missing. In fact, \textit{in-vivo} classification is particularly challenging due to image distortions, specular reflections, occlusions and changing image quality.

Therefore, in this work we analyze deep learning techniques for \textit{in-vivo} hyperspectral laryngeal tumor tissue detection. For this purpose, we evaluate 2D spatial and 3D spatio-spectral CNNs \cite{halicek2017deep, halicek2018optical} in combination with a state-of-the-art Densnet architecture \cite{Huang2017}. Moreover, we propose an efficient network input, based on statistical summaries of the spectral dimension, which reduces the network complexity notably, while achieving competitive results. For evaluation we use an \textit{in-vivo} data set with HSI of the oral cavity or oropharynx of 100 patients, who were examined due to mucous membrane abnormalities in the area of the upper aerodigestive tract.
Summarized, we demonstrate \textit{in-vivo} laryngeal tumor detection using deep learning techniques and HSI.

\section{Methods and Materials} 
\subsection{Data Set}
For evaluation of our methods we use an \textit{in-vivo} data set with HSI image cubes of the oral cavity or oropharynx of patients who were examined due to mucous membrane abnormalities in the area of the upper aerodigestive tract. The full data set consists of 100 patients and was collected at the department of otorhinolaryngology at the university of Bonn. 

Data acquisition was performed in parallel to tissue collection for diagnostic evaluation under general anaesthesia. In this regard, the following experimental setup was utilized. For illumination, a special Polychrome V monochromator (TillPhotonics, Gr\"afelfing, Germany) coupled to a rigid optic using a fiber optic light cable was used. Moreover, an endoscope was fixed in an endoscope holding system (all endoscopy devices: Karl Storz GmbH \& CoKG, Tuttlingen, Germany), to minimize motion artifacts. For image acquisition a monochromatic CCD-camera (AxioCamMRm, Carl Zeiss Microimaging GmbH, G\"ottingen, Germany) was used. The camera was coupled to the endoscope using a dedicated C-mount coupler. 
For each subject, the field of interest of the mucosa was focused during data acquisition and a HS cube was generated with a spectral range from 380 nm to 680 nm in 30 iterative steps of 10 nm. Images were acquired with a spatial camera resolution of $1040\times1388$ pixels. This study was approved by the local ethics committee (\# 176/10 \& 061/13, University of Bonn). 

For data labelling, clinically healthy areas and clinically suspect areas were marked in the recorded images by medical experts. Next, the clinically suspect areas were histopathologically confirmed. For the healthy areas, no tissue extraction was performed, due to ethical reasons. The following Figure \ref{fig:example_images} shows example images with the marked areas. In this work we focus on classifying the marked areas to obtain a well-defined ground truth. While all images contain a marked suspect area, not all contain a marked healthy area. In particular, out of the 100 subjects, 70 subjects have both areas marked and 30 only have a suspect area marked.

\begin{figure}
\centering
\includegraphics[width=0.46\textwidth]{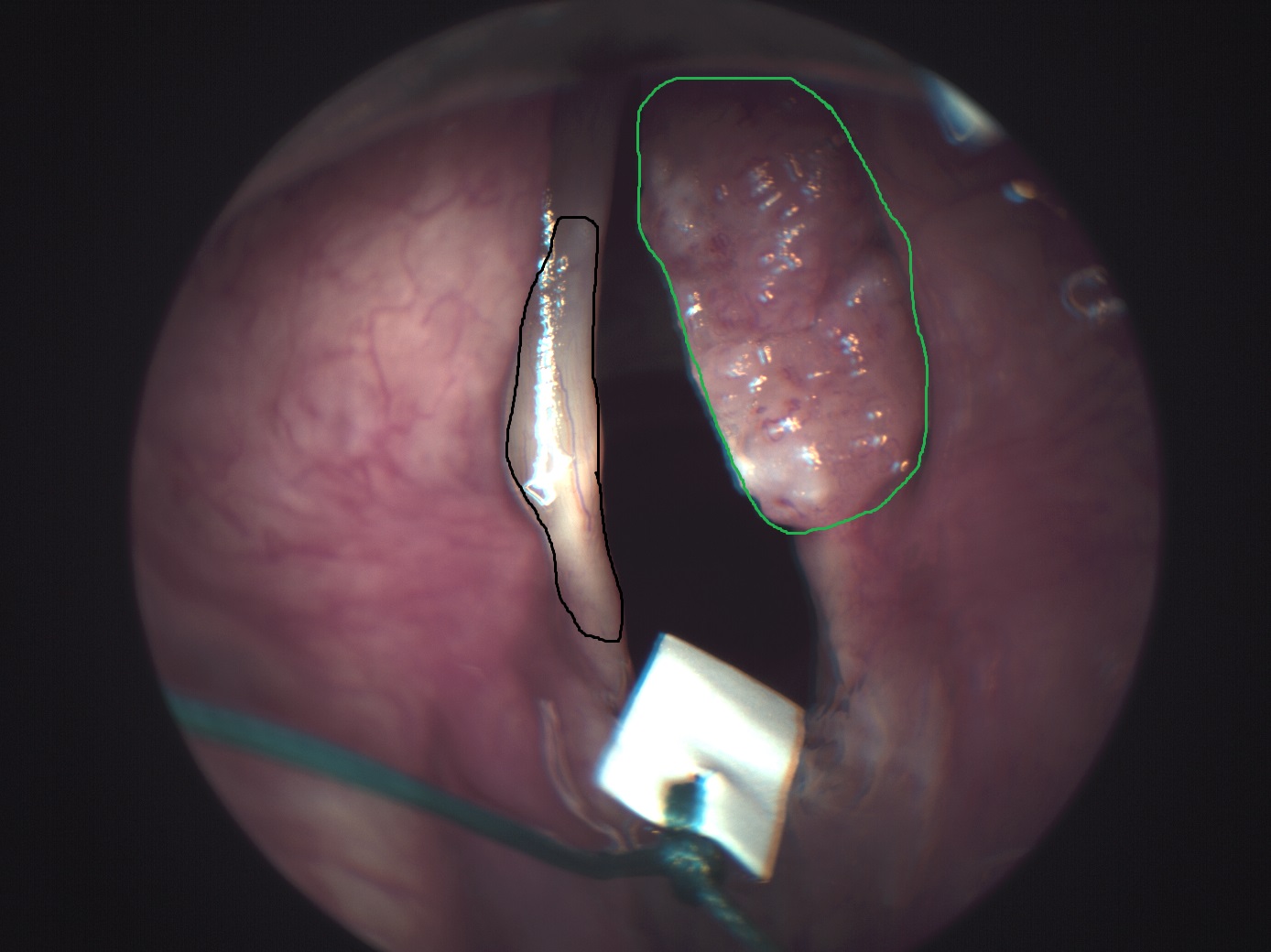}
\includegraphics[width=0.46\textwidth]{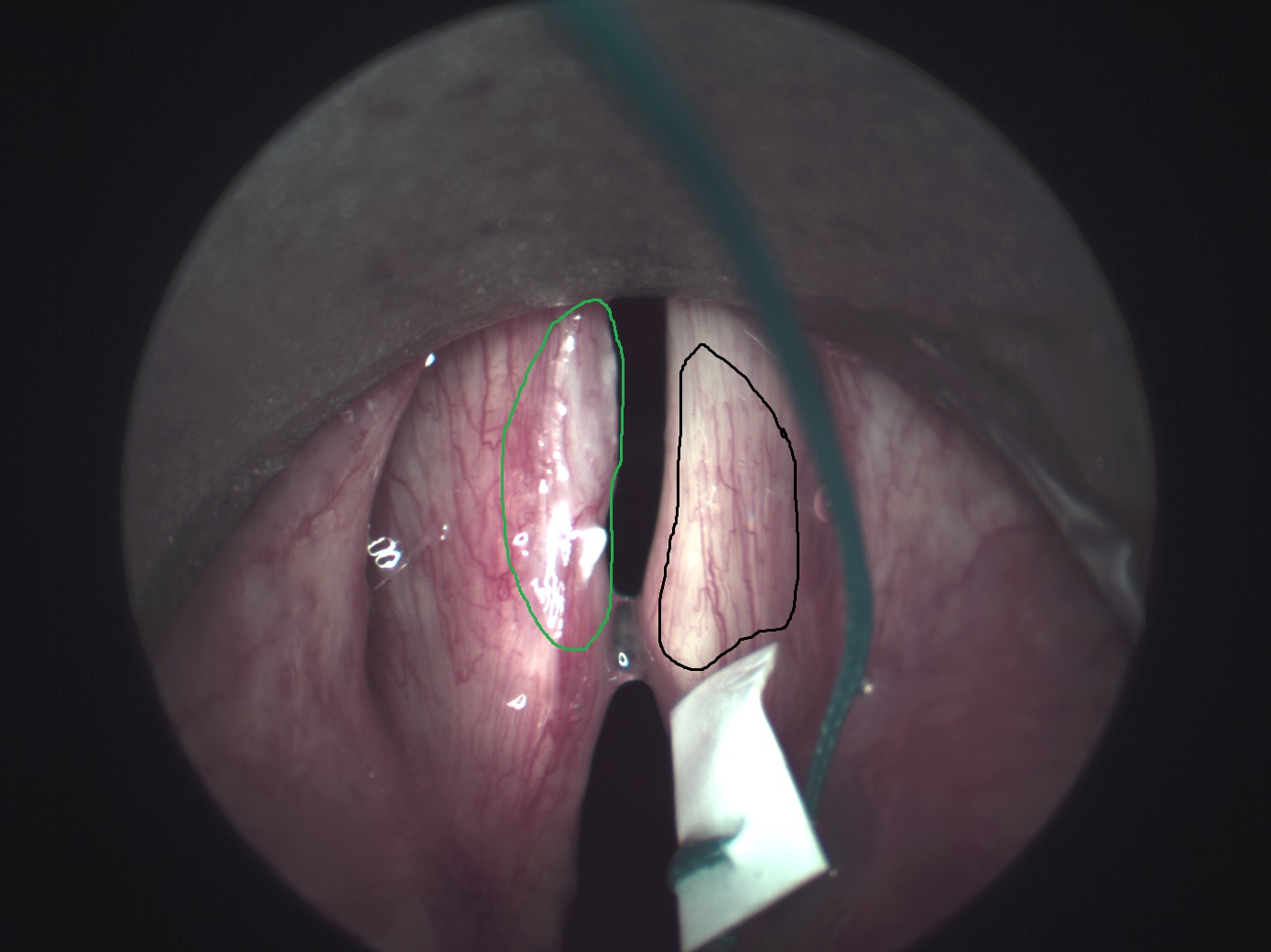}

\caption{Example images of the laryngeal cancer data set shown as an RGB image derived from the HSI cube. The green area segments a region, which is labeled as tumor tissue. The black area refers to a region, which is labeled as healthy tissue.}
\label{fig:example_images}
\end{figure}

Before applying our deep learning methods, we used the following preprocesisng pipeline. First, hyperspectal image stacks were aligend using the ImageJ-implementation of the SIFT-algorithm “Linear Stack Alignment with SIFT” \cite{lowe2004distinctive}. Next, the HSI cubes were filtered using the minimum noise fraction (MNF) \cite{green1988transformation} transformation, as recommended in a previous study on preprocessing of operational HSI of the head and neck area \cite{regeling2016development}. 

To evaluate and train our CNNs, we crop sub-images with a size of $32\times32$ pixels out of the marked areas of the full image. These image crops serve as the input for our CNNs. Note, some pixel regions of the marked areas show extreme specular reflections, which are based on wet surfaces completely reflecting incident light. For this purpose, we excluded image crops, which contain regions with extreme specular reflections, by filtering with an intensity threshold.

\subsection{Deep Learning Methods} 
\begin{figure}
\centering
\includegraphics[width=0.25\textwidth,angle = 90]{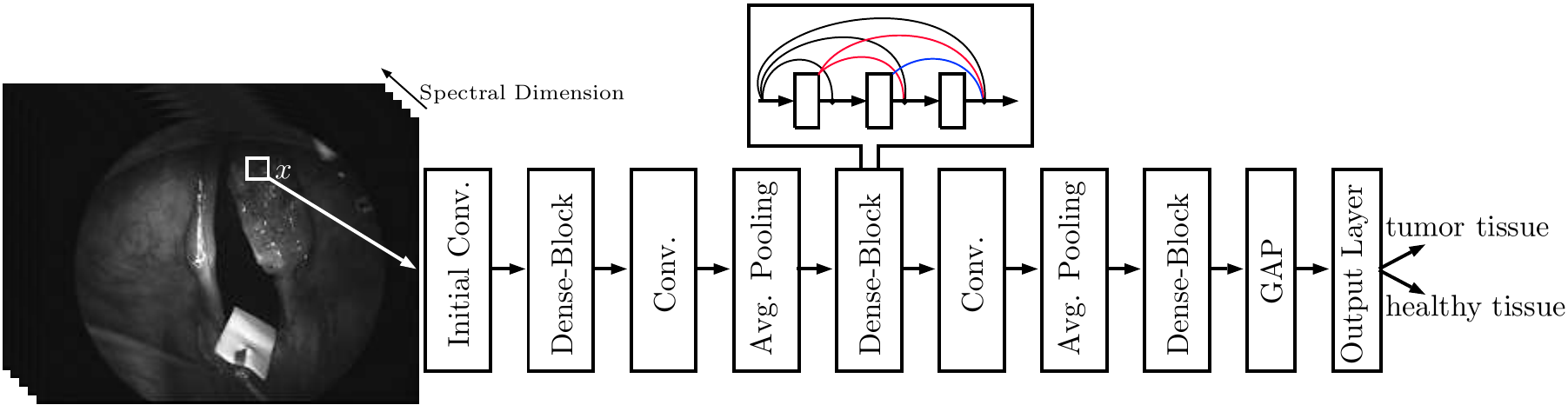}
\caption{Our baseline architecture with three Densenet-Blocks and a global average pooling (GAP) layer before the output layer. The architecture receives a subimage $x$ cropped from the full HSI cube to detect tumor or healthy tissue.
}
\label{fig:model}
\end{figure}

\begin{figure}
\centering
\includegraphics[width=0.8\textwidth]{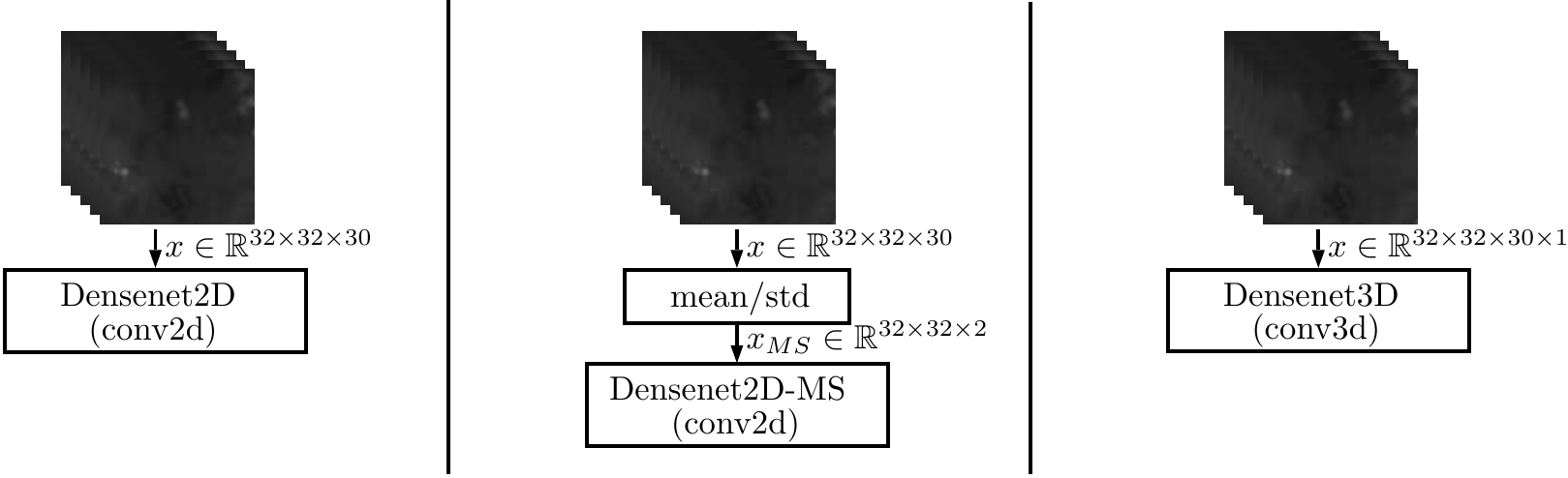}
\caption{Combined with our baseline architecture we evaluate three different methods to learn from the subimages $x$, which are cropped from the full HSI cube. (Left) Using 2D convolutions with the input's channel dimension as spectral dimension. (Middle) Employing 2D convolutions and using the pixel-wise mean and standard deviation as input. (Right) Using 3D spatio-spectral convolutions and treating the data as three dimensional.}
\label{fig:methods}
\end{figure}

\textbf{Models.}
To detect laryngeal tumor tissue our deep learning models receive image crops with a size of $32\times32\times30$ (high-width-spectral dimensions), cropped from the marked areas of the images, together with the corresponding label of tumor or healthy tissue. As a baseline architecture for our CNNs we adopt the idea of of densely connected neural networks (Densenet) \cite{Huang2017}. We use one initial convolutional layer, followed by three Densenet-Blocks, which are connected with transition layers. Before the classification layer we use a global average pooling layer (GAP). The architecture is shown in Figure \ref{fig:model}. For this baseline architecture we evaluate and compare three different methods to learn from the HSI data, shown in Figure \ref{fig:methods}. Hyperparameters are tuned individually for each network. 

First, we use 2D spatial convolutions and stack all spectral dimensions into the input's channel dimension \cite{halicek2017deep}. Hence, the input of our network (Densenet2D) is $32\times32\times30$. Note, 2D convolutions only convolute over the spatial dimensions, thus the spectral dimension is only processed as channel dimension. \\  Second, we use 2D spatial convolutions and we statically summarize the spectral dimension for each pixel. For this purpose, we estimate the pixel-wise mean and standard deviation, thus we obtain one image with mean pixel intensities and one image with the standard deviation of the pixel intensities, which we stack into the input's channel dimension \cite{li20182}. Consequently, the input of our network (Densenet2D-MS) is $32\times32\times2$. \\
Third, we jointly learn from the spectral and spatial dimension by employing 3D convolutions \cite{halicek2018optical}. Thus, the input of our network (Densenet3D) is $32\times32\times30\times1$. \\
\textbf{Training and Evaluation.}
Due to the small data set size, we use 8-fold cross-validation. For each fold, we equally split the data into a test and validation subset. Subsequently, we leverage the validation subset for hyperparameter tuning. For reporting test performances, we average the metrics over all test folds. For data augmentation we use random cropping and random flipping during training. To evaluate the performance for a marked area, we use ordered crops and the predictions of all crops are averaged to obtain one classification. We train our models for 300 iterations with a batch size of 20 and Adam for optimization. To counter the class imbalance, we weight the loss of the individual classes inversely proportional to samples of each class.

\section{Results} 
\begin{figure}
\centering
\includegraphics[width=0.60\textwidth]{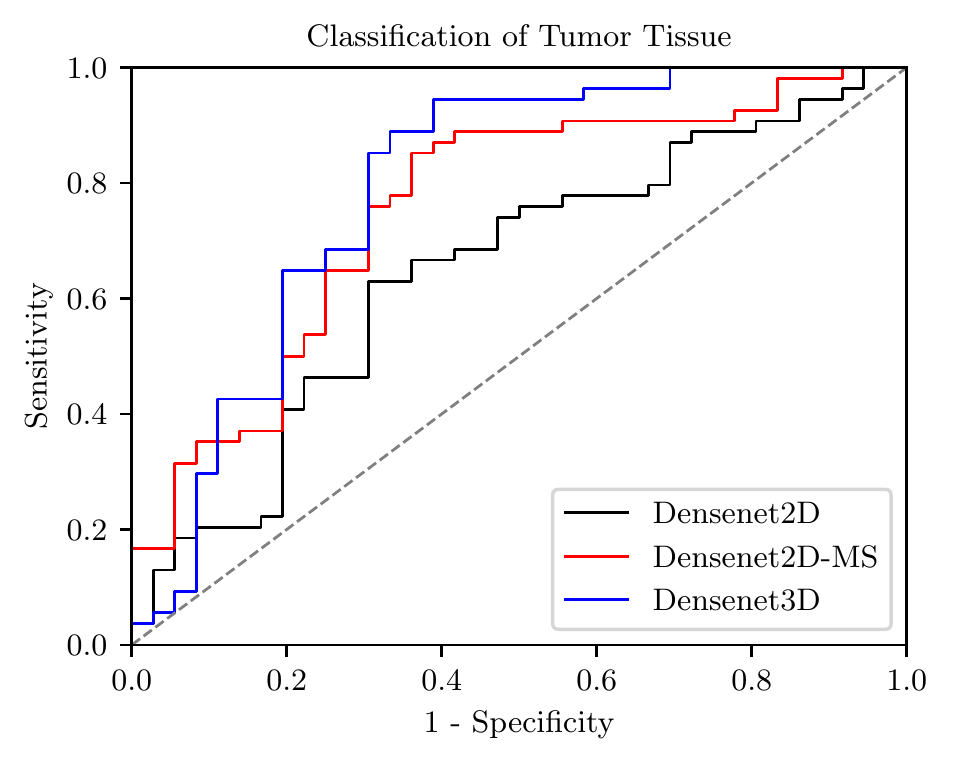}
\caption{ROC curve for the different models shown for the detection of laryngeal tumor tissue.}
\label{fig:ROC}
\end{figure}

The results of all our experiments are shown in Table \ref{tab:All-networks-with metrics}. We consider accuracy, sensitivity, specificity and the F1-Score. For each metric, we report the mean and standard deviation averaged over all cross-validation folds. Densenet3D performs best with a high sensitivity and specificity, followed by Densenet2D-MS. Across all metrics Densenet2D shows the worst performance. In addition, Figure \ref{fig:ROC} shows the receiver operating characteristic (ROC) for the different methods. In this regard, Densenet3D and Densenet2D-MS perform similar for low false positive rates. However, overall Densenet3D clearly shows the best performance.

\begin{table}
 {\caption{Results for all experiments. Sensitivity and specificity are reported with respect to classifying an image as tumor tissue.} \label{tab:All-networks-with metrics}}%
\centering
  {\begin{tabular}{lllll}
  & \bfseries Accuracy & \bfseries Sensitivity & \bfseries Specificity & \bfseries F1-Score  \\ \hline
  Densenet2D & $0.64\pm0.13$ & $0.69\pm0.19$ &  $0.62\pm0.41$ & $0.67\pm0.12$  \\
  Densenet2D-MS &$0.75\pm0.13$ & $0.90\pm0.08$ & $0.54\pm0.24$ & $0.76\pm0.11$  \\
  Densenet3D  & $0.81\pm0.09$ & $0.92\pm0.12$ & $0.65\pm0.21$ & $0.82\pm0.09$ \\
\hline
  \end{tabular}}
\end{table}

\section{Discussion and Conclusion} 
We study deep learning methods for \textit{in-vivo} detection of laryngeal tumor tissue using HSI. Automatic decision support independent of the clinician could greatly increase the chance of an early detection and a successful treatment. For this purpose we investigate the use of CNNs for \textit{in-vivo} HSI. 

Considering a data set of 100 patients and applying 8-fold cross validation, our architecture with 3D convolutions (Densenet3D) performs best with an mean F1-Score of $82\%$. Moreover, our method Densenet3D achieves a high mean sensitivity of $92\%$, while also having a good specficicity of $65\%$. This fits the requirements of a clinical decision support system, which needs a high sensitivity to detect all tumors. In fact, considering the limited training data and the challenges associated with \textit{in-vivo} classification, our results are promising.

Using 2D CNNs and using the input's channel as the spectral dimension turned out to be ineffective. This demonstrates that the 2D CNN does not generalize well if all the potentially redundant spectral information is stacked into the channel dimension. 
Instead, our method with 2D convolutions and using the mean and standard deviation of the pixel intensities along the spectral dimension turned out to be effective. This indicates that reducing the spectral dimension is beneficial for processing with a 2D CNN. 

Moreover, all metrics show a relatively large standard deviation, which is caused by variations across patients. This can be explained by the heterogeneity of the \textit{in-vivo} data set where different view-points, light intensities, or occlusions are present. Thus, robust and generalizable deep learning model development is a challenging task that needs to be solved for automatic clinical decision support with HSI.

In general, our methods have the advantage that once a network is trained, it does not require excised tissue samples and purely operates as an optical assessment method based on HSI. We achieve a high performance on a challenging \textit{in-vivo} dataset which is a promising advancement towards clinical decision support for laryngeal tumor detection.
In conclusion, we demonstrate the ability of deep learning methods to detect laryngeal tumor tissue based on \textit{in-vivo} HSI. Future work could also focus on differentiating between precancerous and cancerous lesions, and segmenting tumor boundaries based on \textit{in-vivo} HSI.

\bibliography{report}   
\bibliographystyle{spiebib}   

\end{document}